# New Recruiter and Jobs: The Largest Enterprise Data Migration at LinkedIn

Xie Lu, Xiaoguang Wang, Xiaoyang Gu[1]

LinkedIn Corporation

# Introduction

In August 2019, we introduced to our members and customers the idea of moving LinkedIn's two core talent products — Jobs and Recruiter — onto a single platform to help talent professionals be even more productive. This single platform is called the New Recruiter & Jobs.

While being a relatively clear and easy sentiment to explain, the updating and migrating of the backend systems was not as straightforward due to the difficulty in migrating their existing data from the legacy database to the new database. To put it plainly, as an engineering team we needed to move our customers from one plane to another while in flight.

The biggest obstacle to change is how to ensure that the data is migrated to a different storage system consistently, at scale, without disrupting business. In order to ensure a smooth transition for all existing customers, we established the following two success criteria for the data migration effort:
- **No Data Discrepancy** - To ensure a seamless transition, customers shall not see any data discrepancies after onboarding to the New Recruiter and Jobs experience. Data discrepancies include not only data loss and data difference but also data duplication and rebirth of deleted data.
- **No Downtime** - LinkedIn offers enterprise products with high reliability and availability for customers all over the world. Recruiter is an essential tool for talent professionals to complete their day-to-day job. Therefore, no downtime, outage, or maintenance shall be introduced for the data migration purpose. It is preferable though to roll out the New Recruiter and Jobs experience at non-peak hours to minimize the disruption on the customers.

In the following, we will discuss the general architecture for a successful data migration and the thought process we follow. Then we expand these ideas to our circumstances and explain in more detail about our specific challenges and solutions. In the Ramp Process section, we explain the inherent difficulties in satisfying our success criteria and describe how we overcome these difficulties and fulfill the success criteria practically.

---

[1] Emails:{xilu,xwang3,xgu}@linkedin.com

# The Anatomy Of A Successful Data Migration

A data migration is a process that replicates data from a source dataset to a destination dataset within a given time frame. A data migration is convergent if and only if the destination dataset converges to the source dataset, and the convergence happens within expected timeframe. A successful data migration requires a data migration system that achieves the convergence between the datasets. In order to build such a data migration system, the engineering process to build the system needs to converge to be bug-free. These are the 3 basic principles we use to assess our data migration:
- Convergence in data
- Convergence in time
- Convergence in engineering process

## Convergence in Data

The following figure demonstrates the overall architecture of a data migration system. In order to successfully migrate data from the legacy database to the new database, there needs to be some important building blocks including dual write components, online/nearline validation and fix components, validation queue, offline validation systems. In the following, we work out the rationale behind these building blocks.

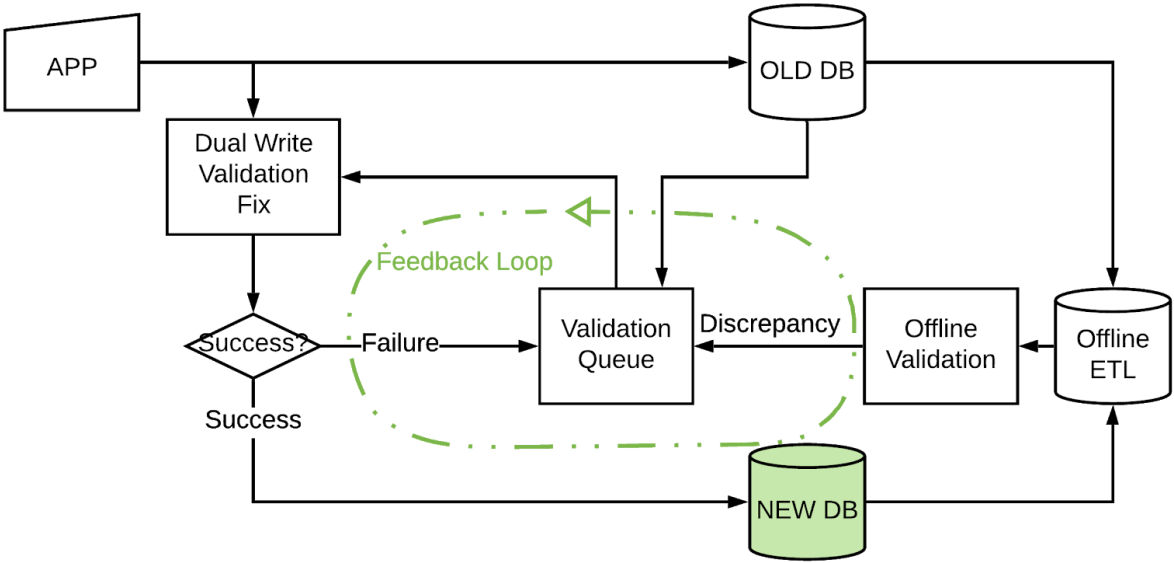

Figure 1: The feedback loop for data convergence

First of all, data needs to be written to the new database so we have to have a mechanism to replicate the writes (dual write) on the legacy database to the new database. It needs to be resilient against failures. So there needs to be a mechanism to detect and re-attempt the failed writes. This determines that the data migration system has to be a closed feedback loop with the ability to *validate* the data and *fix* the data if necessary. There needs to be a mechanism to keep track of the process of validating and fixing data, which is usually a queue. In order to minimize the data broken window, the validation needs to happen as soon as practical after any source of truth data change (legacy database before switch over). Therefore, there needs to be a nearline trigger mechanism to cross-validate the dual writes are successful. Sometimes, these systems or the new database may suffer prolonged outage. To recover from that, the feedback loop needs the ability to validate all the data. Generally this can be done either online or offline while for any database that is sufficiently large, it may become impractical to do full validation online.

Note that the goal of the validation and fix part of the loop is to validate the new database has correct value in comparison to the latest value in the legacy database and the goal is NOT to ensure a particular change is replicated to the new database. It is counterproductive to try to replicate an outdated value. Having the legacy database as the source of truth provides us with a particular kind of idempotency guarantee in that doing validation and fix more than once will not lead to data corruption. This idempotency allows us to repeat validation and fix as needed without worrying about negative impacts and is a key reason the feedback loop can work.

In order for the data to converge, the feedback loop needs to be able to converge. Assume that the system has an availability of $0 < p < 1$ and assume all other things are perfect (bug free and no other failures beyond random unavailability). If the total number of data records is $N$, the total number of validation and fix attempts on individual records needed to fix all data is bounded by $\frac{N}{p}$ and each successive iteration only processes substantially fewer numbers of records. When system availability $p$ is reasonable, the total number of records (including re-attempts) to process is not so much more than the total number of records in the database. (If $p = 0.99$ or $p = 99\%$, we only process a bit over $1.01N$ records or just a bit over 1% more than going over every single piece of data once.) The convergence in the total number of records to process tells us that the feedback loop can converge.

Non-random failures can and do occur. When that happens, a full database validation is often needed. For large scale data migration, offline validation is typically much more scalable. Also, this provides an efficient way to bootstrap the existing data into the data migration feedback loop. Ideally, in a steady state, offline validation is not absolutely required. However, it can always be there to provide metrics and to provide cover for failures in online/nearline systems.

## Convergence in Time

For any update to a piece of data to the legacy database, the *settlement time* for that update in the new database is the earliest time that its corresponding data, that is at least as fresh as the original update itself, becomes available in the new database. Note that we are using clock time here and we do require the clocks to be in reasonable synchronization. Small amount of clock drift here does not make a fundamental difference in the discussion in this section.

For a data migration, *time to converge* for an update to a piece of data is the difference between the *settlement time* of that update in the new database and the update time in the legacy database. At any snapshot of the legacy database, the last update time of the snapshot is the latest update time in that snapshot. The *settlement time* of that snapshot is the latest *settlement time* (in the new database) of all the updates before that snapshot is taken. (Note that in online/nearline settings, this requires certain guarantee of ordering of dual writers.) The *time to converge* for that snapshot is the time difference between the snapshot *settlement time* and the snapshot last update time. The *time to converge* of a time window is the maximum of time to converge of all the snapshots in that time window. When the system is in a steady state, empirically we can use the time to converge of a sufficiently large window as the current *time to converge* of the database.

For data migration, combining migration write operations and the original source of truth database write into the same transaction is generally impractical. So eventual consistency is usually the consistency model for data migration. In this consistency model, the *time to converge* is especially important for the data migration use cases where no down time is *expected*. The asynchronous nature of data migration can and will introduce data races when the source of truth database transitions from the legacy database to the new database potentially leading to broken data. When we switch the source of truth of the database, any updates that have not settled may be lost or end up in corrupted states. This is why any such effort must strive to minimize the time to converge which minimizes the potential data corruption or reduces the required down time. The data migration system is an eventual consistency system with *bounded staleness*. The bound on the staleness is the bound on the potential down time. To have no down time is to minimize the staleness bound to a level that is practically unnoticeable. With proper engineering and change management process in place, it can result in practically no customer perceived down time.

One of the key metrics of the feedback loop is the number of active data points in the feedback loop at steady state (queue length). The processing throughput provided by the system capacity and the queue length determine the time to converge. Note that after the whole feedback loop is running for some time, all the data active in the feedback loop are very recent writes. The maximum age (in the legacy database) of the data point in the loop provides another indicator for the time to converge. Combining these indicators together provides us a practical way to measure the staleness bound for the migrated data in the destination system.

## Convergence in Engineering Process

For everything we described above to actually work in a data migration, the migration system needs to be bug-free. As with any engineering, the building of a data migration system itself is a process. While all engineering projects follow more or less an iterative process before reaching the launch criteria, data migration projects have a unique characteristic of having a well-defined source of truth to compare against as one of the launch criteria. We utilize this to our advantage by building our engineering process around this. The convergent engineering process is illustrated in the flow diagram below. In this process, we continue to monitor the data convergence level resulting from the data migration system. By looking at the amount of data that is not fixed in the expected timeframe, we can determine whether the discrepancy is largely due to random availability failures or more systematic failure due to some bugs in the system.

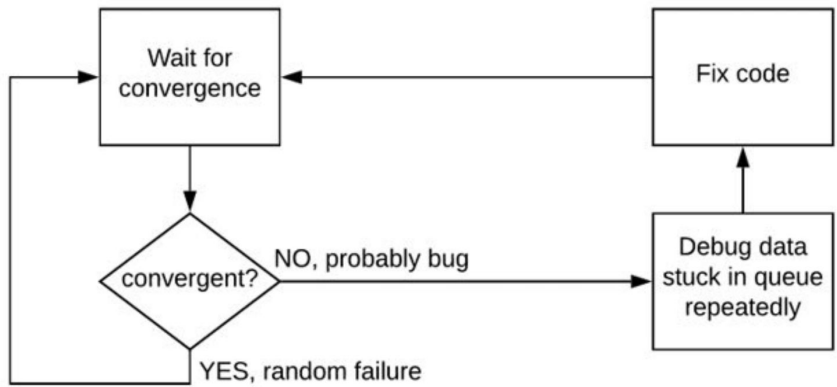

Figure 2: Engineering process convergence

The convergence of this process can be reasoned as follows. We have very good infrastructure availability. If the system is bug-free, the failures should be very rare and should reflect the level of availability. Together with the feedback loop in the data migration system, we should be able to reduce the failure rate to negligible levels (time to converge stay very low), which should also reflect very well on data convergence level. If data convergence does not reach the low level warranted by infrastructure availability in combination with our design, the only possible explanation is there are bugs and we have the actual data samples (stuck in the queue in the feedback loop) to debug the system. Then we fix our code and repeat this process.

The argument so far only considers a static environment where we keep on reducing bugs in the system. In reality, we do not have a static environment. New features are constantly being added. New bugs can be introduced as part of the data migration effort or from other developments. Due to the special nature of the Enterprise business, we are not able to do percentage ramps in quick succession. We have to balance customer need, customer service capacity and take a gradual approach for our ramps and as a result, the entire process lasts over a year. To ensure the true convergence of the process, it is critical throughout all these ramps to remain vigilant, keep a high bar of craftsmanship, improve and follow established processes diligently. In reality, the first few ramps had great outcomes as we designed for, although we did bump into some glitches here and there that required last minute mitigations. Subsequently, we improved our process to ensure smoother operation and later ramps went on smoothly.

We will revisit this Convergent Engineering Process again when we discuss the details of the self-healing property of our system.

## Convergence Metrics

In order to ensure the convergence of the data migration, we need to be able to tell whether the convergence is happening within the desired parameters. In here we list some important metrics to measure:
- Data consistency overall as a general quality indicator
- Data consistency among data not recently updated, which should stay at 100% as an indication that data that should have converged did actually converge
- Active data in the feedback loop -- an indicator of time to converge
- Time in queue -- should remain low

# Design Principles and Choices

Given the Success Criteria we set up, the New Recruiter and Jobs data migration system was designed based on the following principles:

## Full Coverage

Data migration should cover all existing customers regardless of when they are being rolled out to the New Recruiter and Jobs experience. LinkedIn's hiring platform is a multi-tenant environment, there is no difference in ways of migrating different customers' data. This simplifies engineering operations and allows customers to have more flexibility to choose when to adopt the new experience. Data migration should also happen to inactive customers whose data is covered by retention policy so that legacy storage systems and application services can be completely decommissioned after the migration.

## Historical Data Bootstrap

All historical data needs to be backfilled into the new system at a controlled rate by leveraging spare capacity of the system. Bootstrap can take as long as multiple days to complete while it should be a one-time effort that guarantees all historical data are correctly migrated. Bootstrap logic should be shared with the continuous data synchronization wherever possible to establish a consistent pattern of data migration.

## Continuous Data Synchronization

For the data migration to work, all we absolutely need is for the data to converge in the correct time frame when the source of truth switch over happens. In practice, if we start to move large quantities of data during the final days leading up to the switch over, it may create surges in system resource usage leading to potential system instability. We decided to run the data synchronization feedback loop continuously so that the additional stress on the system remains proportional to real traffic and we plan our capacity accordingly. As the data inconsistency is continuously maintained at a very low level, there need not be any last minute scramble to validate or fix large amounts of data and the operation can be much more smooth and predictable. When all underlying infras are operating normally, the final switch over will be a noop besides flipping the switch at a predetermined time.

## Self-Healing Capability

The data migration system should be capable of detecting and reacting to its malfunctions. It should have the ability to examine any data discrepancies caused by system failures and to take appropriate corrections in a reasonable amount of time. The amount of data discrepancy should converge to 0. When all the underlying infrastructures are all operating normally, the time it takes to discover and fix data discrepancy should converge to 0, so that corrections can be made before exposing bad data to customers.

When there are persistent failures in the underlying infrastructure, the above convergence can be violated. Under such circumstances, the source of truth database cannot be switched. However, after the systems recover from the failures, the feedback loop will detect and fix all the resulting data discrepancies and the convergence would resume subsequently.

## One-Way Data Sync

The NRJ is a fully redesigned experience for LinkedIn Recruiter plus many new product features after many rounds of customer research. From the beginning, it was meant to be an upgrade for all customers instead of an option for the customer to choose. With that in mind, it would have been nicer to have the option to go back and forth between two UIs and backend systems. In order to have that level of flexibility however also requires us to build another data sync system that has to work perfectly and is meant to be thrown away. To make it more complicated, the behavior between the two systems are not completely translatable. With all that in mind, the sensible thing to do is to do one-way data sync and make sure it is done really well.

## Full Situational Awareness

Monitoring and alerting is required in every aspect of the migration process. Consistency metrics, scalability metrics, and system health metrics are crucial to the success of a self-healing data migration system. Ultimately all the metrics in place should help answer if data discrepancy has converged to 0 at the time of pulling the ramp trigger. We also introduce some internal alert emails on entity level with more specific information during nearline validation and fix, e.g., any true alarm triggered from offline flow indicating whether nearline systems are working as expected. This is essential for supporting the convergence in process by enabling engineers to investigate and troubleshoot any potential issue in the feedback loop.

# Unique Challenges

## Eventual Consistency

The storage system for New Recruiter and Jobs has evolved from a single shard RDBMS to LinkedIn's homegrown distributed key-value store, making it extremely hard to implement distributed transactions between two systems or maintain constraints over entities. Without having a distributed transaction, eventual consistency is the best the migration system can achieve. So we rely on the nearline and offline feedback loop to double check our data to ensure data consistency.

## Complex Many-to-Many Entity Mapping

New Recruiter and Jobs backend data modelling was significantly different from legacy Recruiter application. This remodelling helps the new hiring platform to be more flexible to incorporate different product use cases, and to avoid putting hard technical limits on its ability to innovate. Therefore, entity mapping from the legacy system to the new system is usually not as straightforward as one-to-one mapping. There are scenarios where one legacy entity maps to multiple new entities and vice versa. These complex entity mappings are very specific to the requirements of the application and there is really no end-to-end technological solution for them. A high level of craftsmanship, attention to detail, uncompromising diligence are required to ensure the code quality in this area to secure the success of the data migration. Adding certain application triggered validation at a layer closer to frontend (shadow read) helps to surface potential data mapping discrepancies. Organized bug bashes where a group of engineers inspect the data shown in legacy and new UI is still the tried and true process to uphold the quality bar.

## Entity Dependency Graph

Entities to migrate are usually not independent. For example, a candidate can only be added to a project after the project is created; a candidate can only be moved to a state after the state is created in the hiring pipeline. Therefore, the order of different entities data synchronization must strictly follow the dependency graph.

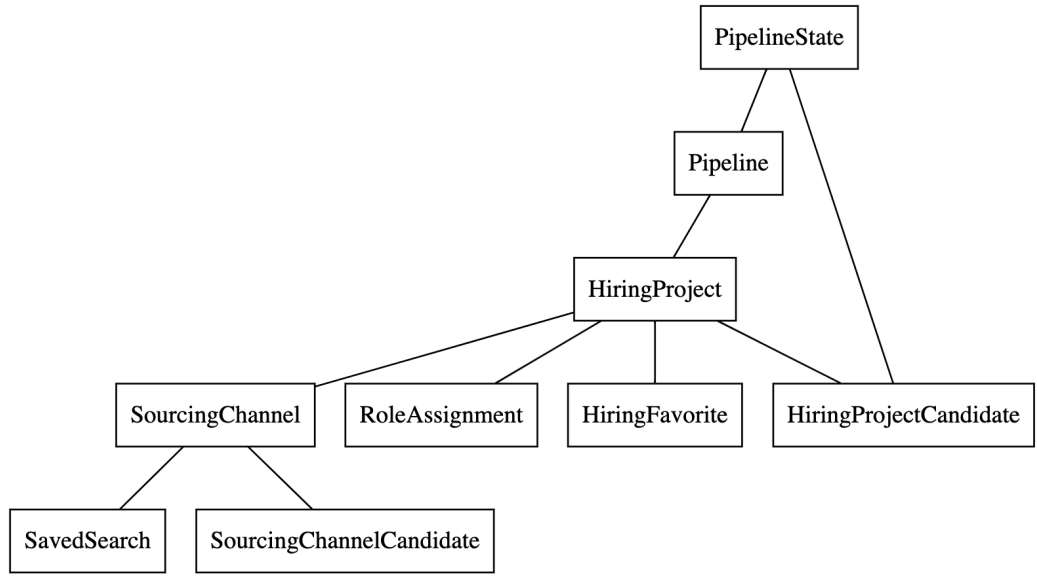

Figure 3: Sample entity dependency graph

## Bulk Operations

LinkedIn Recruiter allows customers to perform bulk operations including candidate imports from spreadsheet, seat transfers, removing a state from hiring pipeline, etc. These bulk actions usually end up with a bursty write traffic to the source database and create a challenge for data migration to keep time to converge below minutes. These are situations where there is really no practical technological solution. In order to ensure the success of our customer ramps, we decided to coordinate with the customer services team to disable all large bulk operations for the three days leading up to planned customer ramps.

# System Architecture

In the New Recruiter and Jobs data migration system, we implement the aforementioned feedback loop into the system with the following key parts:
- Offline bulk loads for historical data bootstrap
- Online dual writes for continuous data synchronization
- Nearline change stream verification for self-healing capability driven by data change capture
- Online shadow read verification for self-healing capability driven by user interaction
- Offline bulk verification for self-healing capability with full coverage but some delay

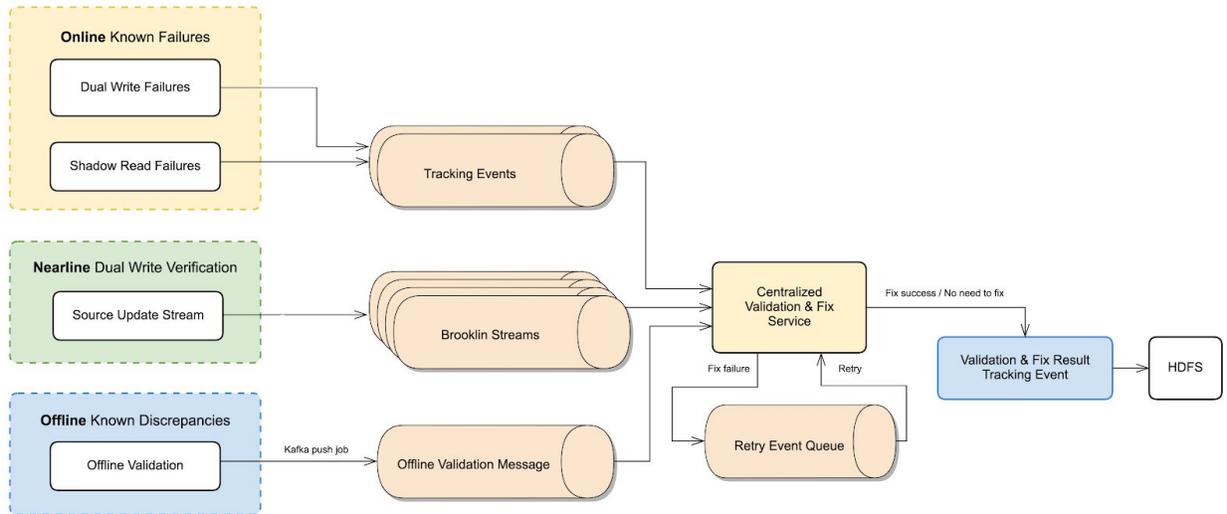

Figure 4: Migration system architecture overview

## Offline Bulk Loads

LinkedIn's ETL infrastructure allows data ingestion from online data storage to HDFS for offline processing. It applies incremental data changes on top of a database snapshot and publishes merged incremental snapshots every 8 to 12 hours.

For each type of entity to migrate, the historical data bootstrap pipeline compares the ETL copy of the source data and target data with predefined entity mapping logic. The bootstrap workflow runs Spark and Pig jobs on Hadoop to prepare the backfill patch for further processing. Due to the entity dependency graph, multiple rounds of backfill are required and they have to happen in correct order.

For entities that contain billions of records to migrate and require only idempotent operations, bulk loading data from offline directly into the target database is preferred. Bulk jobs give priority to live traffic to reduce impact on customers, and they run at the spare capacity of the shared multi-tenant database cluster. Note that certain records backfilled in this way might need further corrections due to the level of staleness of the ETL data.

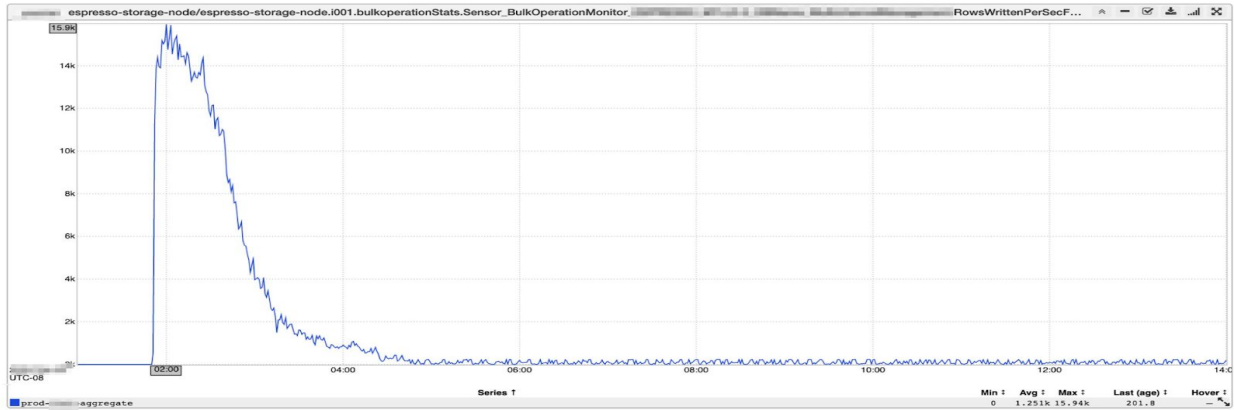

Figure 5: QPS for bulk loading data into target database can be as large as 15.9K

Other entities are fed into the online system for backfill via Kafka push job from offline pipeline. Online system is able to get a latest copy of the source data and write it to the target database at a controlled rate. Similar to offline bulk jobs, the rate limiter is capacity-aware and always gives higher priority to live traffic to reduce impact on customers. The verification and fix logic is shared with other online and nearline flows so that different triggers converge to one consistent way of migrating data.

## Online Dual Writes

For continuous data synchronization, any update in the legacy Recruiter application needs to be dual written to the new Recruiter and Jobs platform. Due to the high complexity of implementing distributed transactions across two different separate application stacks with different storage solutions, writing to the new platform only happens asynchronously upon a successful commit to the legacy system.

Online dual writes are not meant to impact customers so the legacy Recruiter application should be agnostic to the result of dual writes. At the same time, any failures during dual writes need to be tracked and fed into a self-healing queue to trigger validation and fixes.

Despite the fact that online dual writes to two systems may cause temporary data inconsistencies under certain circumstances, it ensures that in most scenarios different types of entities are written to the new platform in the desired order following the entity dependency graph almost instantaneously after the initial write to the legacy system.

## Nearline Stream Verification

Brooklin is the default change-capture service for many data sources at LinkedIn. Any update to the source database will be streamed via Brooklin. This nearline processing flow provides an additional opportunity for the migration system to systematically validate whether the source update in the legacy system is correctly reflected in the new database. Any validation failures during Brooklin stream processing are fed into the self-healing queue for corrections.

The nearline stream validation success rate is closely monitored as an indicator of online dual writes reliability. It is a critical factor to determine whether the time-to-converge is close to zero, i.e., whether last-minute changes in the legacy database can be replicated in the target database before flipping the final switch.

## Online Shadow Read Verification

Nearline stream verification is triggered by every successful write to the legacy Recruiter application while online shadow read flow verifies data consistency upon every successful read. As the name suggests, shadow read flow attempts to read the same data that is expected from the target database, compares the result with the source data, and evaluates if discrepancies exist. Additionally, as we discussed earlier in complex many-to-many entity mapping, the shadow read tends to happen close to API level, which provides additional value in verifying the correctness of entity-mapping and even business logic in the new system.

Shadow read is driven by real user interactions with the system and thus it works like a dry run of switching the data source. Shadow read gives the closest to reality metric about data consistency as if the user were migrated to the new platform.

Similar to online dual writes, shadow reads happen asynchronously upon a successful read from the legacy system. Shadow read results shall not impact user behavior while any known discrepancies detected via shadow reads should be tracked via Kafka events and fed into a self-healing queue for corrections.

## Offline Bulk Verification

Offline bulk load workflows are repurposed to an offline bulk validation job that runs upon every new database incremental ETL snapshot is published after the initial bootstrap. The offline bulk verification flow scans through all records in the legacy database that are updated 24 hours before the job run time and detects discrepancies in the target database. Discrepancies are sent to the self-healing queue for corrections via Kafka push job. It leverages the same validation and fix pipeline that is shared among online, nearline, and offline verification flows.

## Self-Healing Queue

Self-healing queue is an essential part of the feedback loop that enqueues events for data consistency validation and only dequeues events if either the validation is successful or otherwise correction has been applied successfully. The self-healing queue size should diminish over time and remain at very small size during normal operations. If there are events stuck in the queue for an extended amount of time, it indicates a defect that needs immediate action. Any complex errors that cannot be fixed in the feedback loop in a reasonable number of attempts are surfaced to the development team automatically for manual review. The bug fixes are added to the queue processors as needed to achieve the engineering convergence to a bug-free system.

The queue is implemented with Quantum library which provides event scheduling and execution on top of LinkedIn's homegrown distributed key-value store. The queue processors are capacity-aware and rate limited to prevent negative impact on the online system. System health metrics including event counting metrics and latency metrics have been set up for every type of data entity to monitor the self-healing queue at a high level of granularity. Counting metrics include events enqueue count, dequeue count, validation success/failure count, fix success/failure count, and retry count. Latency metrics include events in queue latency, validation and fix latency, and total pipeline latency.

## Ramp Process

By the nature of the data migration system, the legacy system and the new system is a dependent system connected via network. We can model the data consistency problem between the two systems as the consensus problem in distributed systems. Before the customer ramp, all read operations only read from the legacy system (and databases). After the ramp, all read operations only read from the new system (and databases). It means that in this model, the network is practically always partitioned between the legacy and the new system on the read path.

With such a bi-partite distributed system, at most one of the two sides can form a quorum on its own[2]. By the CAP theorem[3], Consistency and Availability cannot be achieved together across the switch over when we artificially introduce the partitioned network[4]. This is a hard limit that we cannot cross unless we choose to either build the new system to be able to read from both new and legacy databases or make all writes in the legacy system transactional across the legacy and new databases, which are extremely costly, impractical (plus the two systems are not exactly functionally equivalent), and potentially negatively impact system availability due to increased system interdependency. Therefore, no matter how robust our system is, we have to trade off between data consistency and availability during the time when the distributed consensus is really needed, namely, when we switch over to the new system.

Between the two important aspects impacting customer experience -- consistency and availability, we choose consistency and introduce a small window of unavailability to allow data sync to settle in the new system so that we do not need a quorum read to get consistent data. The best we can do is to make this unavailability window as brief as practical. Subsequent write operations done from the new system does not require any participation of the legacy system and any subsequent read operations can be local to the new system.

---

[2] Our system is not a quorum based system. In abstract terms, however, our system can not perform fundamentally better than a quorum based distributed system so all the arguments here will stand.

[3] Among Consistency, Availability, Partition Tolerance, one cannot achieve more than two out of three in a distributed system.

[4] One can argue that before the switch over all users are on the legacy side of the network and after the switch over all users are on the new side of the network. In reality, user actions and related data operations take time to materialize and replicate, so the operations on two sides of the network will have a window of overlap.

In order to minimize unavailability window and minimize the customer expose to it, we established and meticulously followed the following processes:
- Monitor all infrastructure systems to ensure that all dependent systems are operating normally and performing at expected level. Monitor and check all offline and online data consistency metrics to ensure all data migration systems are operating normally and performing at expected level.
- Run data migration continuously so that all the data is kept consistent at all times as long as the underlying infrastructures are operating normally. This system is designed to operate continuously and it avoids last moment scrambles to get large amounts of data into consistency before a deadline.
- Stop all bulk operations three days before the planned ramp to ensure consistent low data synchronization latency.
- Schedule the ramps on Sunday mornings to further minimize the amount of data changes in the system. This also gave us the benefit of having nothing else to deal with in case any issues arise.
- Log out customer sessions right before ramp time to force them into the new UI, which introduces a small window of unavailability.
- Aside from the regular monitoring of data quality, we set up ramp clearance meetings four days before the planned ramps to go over data consistency reports. We also set up ramp criteria clearance meetings much earlier to ensure the list of included customers qualify all the criteria we set up.

The data migration system does not exist in isolation. Instead, the system lives in a dynamic environment where things do change. Establishing and following the processes meticulously is key to detect and address potential problems early. The consistency in adhering to the processes over time is instrumental to our continued success.

## Conclusions

The New Recruiter and Jobs data migration is the one of the largest and most complicated enterprise data migration challenges at LinkedIn. In order to ensure the success of the data migration and customer satisfaction, we formulated the Convergence Principles for data migration, designed for success by following a principled approach, held a high bar of craftsmanship, moved fast with caution, and established and followed necessary engineering processes meticulously. We were able to reach a steady state data consistency rate of better than 99.999% (most often 100%) and bring down the steady state potential broken data window down to no more than minutes (limited only by infrastructure SLA). The biggest obstacle to massively roll out the New Recruiter and Jobs experience has been removed.

With the majority of our customers already ramped to using NRJ, we have not received any customer reports on loss of availability due to the ramp process nor have we received any reports on broken data introduced by the data migration system introduced here.

We find the Convergence Principles formulated above to be particularly illuminating during our design and development. It enabled us to systematically analyze what needs to be built and to what spec they need to be built in order for us to be able to achieve the quality target for data migration. The principles are broadly applicable for all data migration efforts and we hope other efforts can benefit from us sharing our experience here.

## Acknowledgements

This has been a massive initiative for LinkedIn and is a result of several months of herculean efforts from engineers in LinkedIn Talent Solutions organization.

Special thanks to all the developers who made this possible: Sonya Liang, Si Chang, Matthew Deng, April Dai, Shidong Yan, Shuya Wang, Wayne Lu, Bin Gao, Yizhuo Chu, Fan Zhang, Mikhail Kreytser, Yangmei Zheng, Jacob Lin, Andrew Luo, Bo Pan, Chunnan Yao, Wenxin Xing, Cuisheng Yao, Bo Yao, Yefei Wang, and Roman Chang, SREs: Suchet Chachra, Matthew Gyurgyik, Allan Carhart, Mykyta Gubenko, Payman Darabi, Ethan Sunyin, and AFURE Oyibo, cross functional partners: Arya Choudhury, Gautam Dogra, and Jonathan Pohl, and leadership: Aarathi Vidyasagar, Rahul Sule, Shen Shen, Lei Ni, Dan Reid, Michel Moriniaux, and Ajoy Chattopadhyay.

Special thanks to Jingwei Wu and Ketan Thakkar for reviewing and providing valuable feedback.